\documentclass[twocolumn, showpacs,preprintnumbers,amsmath,amssymb,pra]{revtex4}
\usepackage[ansinew]{inputenc}
\usepackage{graphicx}
\usepackage{pst-all}
\usepackage{color}
\usepackage{dcolumn}
\usepackage{amsmath}
\usepackage{amsthm}
\usepackage{bm}
\usepackage{layout}
\usepackage{float}
\usepackage{txfonts}
\usepackage{amsfonts}
\usepackage{amssymb}%
\usepackage[ansinew]{inputenc}
\usepackage{graphicx}
\usepackage{amsmath}
\usepackage{amsthm}
\usepackage{bm}
\usepackage{layout}
\usepackage{float}
\usepackage{amsfonts}
\usepackage{amssymb}
\usepackage{array}
\usepackage{multirow,bigdelim}
\usepackage{slashbox}
\usepackage{enumitem}
\usepackage{graphicx}
\usepackage{dcolumn}
\usepackage{bm}
\usepackage{amssymb}
\usepackage{amsfonts}
\usepackage{graphicx}
\usepackage{dcolumn}
\usepackage{bm}

\usepackage{times,epsfig,amssymb,amsmath}

\usepackage{bm}
\usepackage{times,epsfig,amssymb,amsmath}
\usepackage{graphicx}

\begin{document}

\bibliographystyle{apsrevowner}

\title{Sudden Change of Quantum Discord under Single Qubit Noise }
\author{Li-Xing Jia,$^{1,2}$ Bo Li$^2$,
R.-H. Yue,$^1$  Heng Fan$^2$\footnote{hfan@iphy.ac.cn}}%
\affiliation{%
$^1$Faculty of Science, Ningbo University, Ningbo 315211, China\\
$^2$Beijing National Laboratory for Condensed Matter Physics,
Institute of Physics,
Chinese Academy of Sciences,
Beijing 100190, China}

\date{\today}

\begin{abstract}
We show that the sudden change of quantum correlation  can occur even when only one part of the composite entangled state is exposed to a noisy environment.
Our results are illustrated through the action of different noisy environments individually on a single qubit of quantum system. Composite noise on the whole of the quantum system is thus not the necessarily  condition
for the occurrence of sudden transition for quantum correlation.
\end{abstract}


\maketitle

\section{Introduction}

Quantum entanglement, a kind of nonclassical correlation in quantum world,
is a fundamental concept of quantum mechanics \cite{Naturwissenschaften.23.807-812,PhysRev.47.777}.
It is well accepted that entanglement plays a crucial role and is the
invaluable resource in quantum computation and quantum information \cite{RevModPhys.81.865,Nielsen}.
Recently, it is realized that entanglement is not the only aspect of quantum correlations,
and the nonclassical correlations other than entanglement may also
play fundamental roles in quantum information processing \cite{PhysRevLett.104.080501,PhysRevLett.88.017901,RevModPhys.84.1655}.
Among several measures of quantum correlations,
the so called quantum discord, introduced by  Olliver and Zurek \cite{PhysRevLett.88.017901}
and also by  Henderson and  Vedral  \cite{J.Phys.A.Math.Gen.34.6899}, has been receiving a great deal of attentions
\cite{PhysRevA.67.012320,PhysRevA.71.062307,PhysRevA.76.032327,PhysRevA.77.042303,PhysRevA.80.024103,
PhysRevA.80.022108,PhysRevA.80.044102,PhysRevA.80.052304,PhysRevA.81.042105,PhysRevA.84.042313,
PhysRevA.83.022321,PhysRevA.86.032110,PhysRevLett.100.050502,PhysRevLett.100.090502,PhysRevLett.102.250503,
PhysRevLett.102.100402,PhysRevLett.105.190502,PhysRevLett.104.200401,PhysRevA.85.032318,PhysRevA.86.012312,arXiv:1208.5705}.

Quantum discord, as an important supplementary to quantum entanglement, is found to be
present in the deterministic quantum computation with one qubit (DQC1) while the entanglement
is vanishing \cite{PhysRevLett.100.050502,PhysRevLett.101.200501}. On the other hand, quantum discord is similar as entanglement, for example,
it can be considered as a resource in some quantum information processing protocols \cite{PhysRevLett.107.080401,PhysRevA.85.022328}.
Quantum discord can also be related directly with quantum entanglement \cite{PhysRevLett.106.160401,PhysRevA.69.022309}.
One fundamental point is perhaps that quantum discord coincides quantum entanglement for pure state.
However, besides different roles in DQC1, there are some other fundamental points which are different for quantum discord
and quantum entanglement.
% For example, mixed state contains more quantum discord than that of pure state \cite{winterorbruss},
%while pure state can contain maximal entanglement for a system.
One of the key differences is that quantum entanglement is non-increasing under
the local operations and classical communication (LOCC), while quantum discord can increase
under a local quantum channel acting on a single side of the studied
bipartite state \cite{PhysRevLett.107.170502,PhysRevA.85.032102}.
This is surprising since it is generally believed that quantum correlations can only decrease generally
under local quantum operations even classical communication is allowed.
%This is the case for entanglement, i.e., entanglement does not increase under LOCC.
Here let us note that various quantum correlations, including quantum discord
and quantum entanglement, are invariant under local unitary operations by definition.

To be more explicit, entanglement can increase only when coherent operations are applied.
This is similar as in classical case, we know that classical correlation can increase when
classical communication is allowed, which apparently involves two parties.
In comparison, quantum discord which describes the quantumness of
correlations, can increase under one-side local quantum operations.
In this paper, we will add more evidences to show that quantum discord may possess
some properties under a single local quantum channel.

We know that contrary to the entanglement sudden death (ESD) \cite{PhysRevLett.93.140404,PhysRevLett.97.140403,arXiv:12103216v2},
the behaviors of quantum discord under the Markovian environments decays exponentially and disappears  asymptotically\cite{PhysRevA.80.024103,PhysRevA.81.052318}.
But the investigation recently shows that the decay rates of quantum correlation
may have sudden changes \cite{PhysRevA.80.044102,PhysRevLett.104.200401} under composite noises.
We already find some evidences showing that one side quantum channel
is already enough for the occurrence of some phenomena for quantum discord ,
one may wonder whether the discord sudden change can occur when only one qubit of quantum system is
subjected to a noisy environment while leaving the other subsystem free of noise?
In this article, we investigate the dynamics of quantum discord of two qubit X shape state with only one particle exposed to  noise.
Our results show that composite noises are not necessary for the sudden change of quantum correlation, a single side
of quantum channel is enough.

\section{Classical and Quantum correlations of X shape states}

It is widely accepted that the total correlation of a bipartite system ~$\rho_{AB}$ is measured by the quantum mutual information defined as
\begin{equation*}
\mathcal{I}(\rho_{AB}) = S(\rho_{A}) + S(\rho_{B}) - S(\rho_{AB}),
\end{equation*}
where ~$\rho_{A}$ and ~$\rho_{B}$ are the reduced density matrices of ~$\rho_{AB}$, and ~$S(\rho) = -\text{Tr}\{\rho \text{log}_2 \rho\}$ is the von Neumann entropy. Classical correlation\cite{J.Phys.A.Math.Gen.34.6899} is defined as

\begin{equation*}
\mathcal{C}(\rho_{AB}) = \max_{B_i^\dagger B_i} S(\rho_{A}) - \sum_i p_i S(\rho_{A}^i),
\end{equation*}
where ~$B_i^\dagger B_i$ is a POVM performed on the subsystem B,  ~$p_i = \text{Tr}_{AB} (B_i \rho_{AB} B_i^\dagger)$, and ~$\rho_{A}^i = \text{Tr}_{B} (B_i \rho_{AB} B_i^\dagger)/p_i$ is the postmeasurement state of A after obtaining the outcome on B.
Then quantum discord which quantifies the quantum correlation is given by

\begin{equation*}
\mathcal{Q}(\rho_{AB}) = \mathcal{I}(\rho_{AB})-\mathcal{C}(\rho_{AB}).
\end{equation*}

Consider the following Bell-diagonal states \cite{PhysRevA.77.042303}
\begin{equation}
\rho = \frac{1}{4}\Big(I \otimes I +\sum_{j=1}^3 c_j\sigma_j\otimes \sigma_j\Big),
\label{Eq:rho}
\end{equation}
where ~$I$ is the identity operator on the subsystem, $\sigma_j, j=1,2,3$, are the Pauli operators,
$c_j\in\mathbb{R}$ and such that the eigenvalues of ~$\rho$  satisfying ~$ \lambda_i \in [0,1]$. The states in Eq. (\ref{Eq:rho}) represents a considerable class of states including the Werner states $(|c_1|=|c_2|=|c_3|=c)$ and Bell $(|c_1|=|c_2|=|c_3|=1)$ basis states.
The mutual information and classical correlation of the state ~$\rho$  in Eq. (\ref{Eq:rho}) are given by \cite{PhysRevA.77.042303}
\begin{eqnarray*}
\mathcal{I}(\rho)&=&2+\sum_{l=0}^3\lambda_l log_2 \lambda_l,\label{eq:mutual}\\
\mathcal{C}(\rho) &=& \frac{1-c}{2} log_2 (1-c) + \frac{1+c}{2} log_2 (1+c),\label{classicalcorrelation}
\end{eqnarray*}
where $c = \max \{|c_1|,|c_2|,|c_3|\}$, and then the quantum discord of state ~$\rho$ is given as
\begin{eqnarray*}
&& \mathcal{Q(\rho)} = \mathcal{I(\rho)}-\mathcal{C(\rho)}\nonumber\\
            &=& 2+\sum_{i=1}^4 \lambda_i log_2 \lambda_i-\frac{1-c}{2} log_2 (1-c) - \frac{1+c}{2} log_2 (1+c).
\end{eqnarray*}

A generalization of quantum discord from Bell-diagonal states to a class of X shape state is given in  Ref. \cite{PhysRevA.83.022321}  recently
with state in form,
\begin{equation}
\rho'  =  \frac{1}{4}\Big(I \otimes I + \mathbf{r}\cdot \sigma \otimes I + I \otimes \mathbf{s} \cdot \sigma + \sum_{i=1}^3 c_i \sigma_i \otimes \sigma_i \Big)\label{eq:xstates4}
\end{equation}
where   $\mathbf{r}=(0,0,r),\mathbf{s}=(0,0,s)$, one can find that  ~$\rho'$ reduces to  ~$\rho$ when  ~$r = s = 0$.
The mutual information and classical correlation of state  ~$\rho'$ are given by
\begin{eqnarray*}
\mathcal{I}(\rho') &=& S(\rho '_A) + S(\rho'_B)-S(\rho'),\\
\mathcal{C}(\rho') &=& S(\rho_A') - \min \{S_1, S_2, S_3\}.
\label{liboresult}
\end{eqnarray*}
Here ~$S_1,S_2,S_3$ are shown in \cite{PhysRevA.83.022321}, and  ~$f(t)$ is defined as ~$f(t)=-\frac{1+t}{2}log_2(1+t)-\frac{1-t}{2}log_2(1-t)$.
Then quantum discord, the quantum correlation of state ~$\rho'$, is given by
\begin{eqnarray*}
\mathcal{Q(\rho')} &=& \mathcal{I(\rho')}-\mathcal{C(\rho')}\nonumber\\
            &=& S(\rho'_B)-S(\rho')+\min\{S_1,S_2,S_3\}.
\end{eqnarray*}

It is difficult to calculate quantum discord in general case since the optimization
should be taken. However, the analytical expression of quantum discord
about Bell-diagonal states is available \cite{PhysRevA.77.042303} which provides
a convenient method in studying the dynamics of quantum correlation in case the
studied states satisfy this special form \cite{PhysRevA.80.044102,PhysRevLett.104.200401}.
In this paper, based on an analytical expression of quantum correlation
which generalize the discord from Bell-diagonal states
to a class of X shape states \cite{PhysRevA.83.022321},
we can investigate dynamics of quantum discord with more kinds of quantum noises.
In particular, the analytical discord can be found for cases with one side quantum channel.
As a result, we can study three different kinds of quantum channels,
amplitude, dephasing, and depolarizing which act on the first qubit of a class of two-qubit X states.
We next consider those three quantum channels respectively.

\section{amplitude noise}\label{amplnoise}
Amplitude damping or amplitude noise which is used to characterize  spontaneous emission describes the energy dissipation from a quantum system .The Kraus operators for a single qubit are given by \cite{PhysRevLett.93.140404}
\begin{equation*}
E_{0}=
\begin{pmatrix}
\eta & 0  \\  0 &  1
\end{pmatrix},
E_{1}=
\begin{pmatrix}
0 & 0  \\  \sqrt{1-\eta^{2}} &  0
\end{pmatrix},
\end{equation*}
where ~$\eta = e^{-\frac{\tau t}{2}}$ and ~$\tau$ is the amplitude decay rate, $t$ is time.
We consider the case that the first qubit is through this quantum channel.
So the Kraus operators for the whole system, with amplitude noise acting only on the first qubit,
are given by
\begin{eqnarray*}
K_{1a}&=&
\begin{pmatrix}
\eta & 0  \\  0 &  1
\end{pmatrix}\otimes
\begin{pmatrix}
1 & 0  \\  0 &  1
\end{pmatrix}, \nonumber \\
& & \\
K_{2a}&=&
\begin{pmatrix}
0 & 0  \\  \sqrt{1-\eta^{2}} &  0
\end{pmatrix}\otimes
\begin{pmatrix}
1 & 0  \\  0 &  1
\end{pmatrix} .\nonumber
\end{eqnarray*}

\begin{figure}
  % Requires \usepackage{graphicx}
  \includegraphics[width=0.5\textwidth]{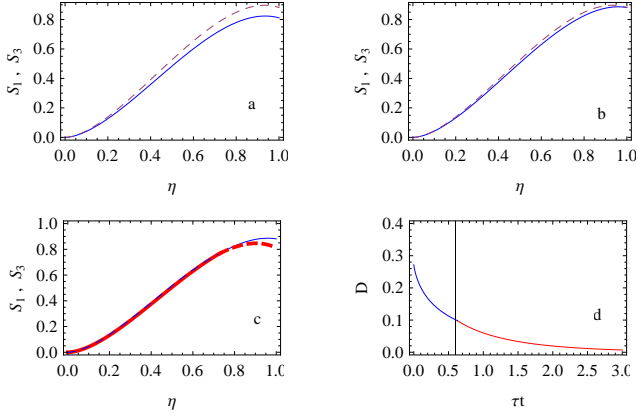}\\
  \caption{~$S_1$(dashed line)\ \ and\ \ $S_3$(solid line)\ \ of\ \ Eq. (\ref{s1amplitude}), \ \ \ \ Eq. (\ref{s3amplitude}): (a) $c_2 = 0.5, c_3 = 0.4$,  ~$S_3 < S_1$ when $\eta \in (0,1)$; (b) $c_2 = c_3 = 0.4$, ~$S_3 < S_1$ when $0< \eta <1$; (c) $c_1 = 0.1,c_2=0.4,c_3=0.5$, ~$S_3 < S_1$ when ~$\eta \in (0,0.73)$   and  ~$S_1 < S_3$ when ~$\eta \in (0.73,1)$; (d) Quantum discord of \ $\varepsilon _a ( \rho )$ with the same parameters as in situation (c), the decay rate of quantum discord sudden changes at ~$\tau t =0.63$.}\label{amplitude4}
\end{figure}

Let $\varepsilon(\cdot)$ represents the operator of the noise environemt.With the time-dependent Kraus operator matrix of amplitude noise acting on the first qubit of state $\rho$, we have
\begin{widetext}
\begin{eqnarray*}
\varepsilon _a ( \rho )  &=& K_{1a} \rho K_{1a}^\dagger + K_{2a}\rho K_{2a}^\dagger \nonumber \\
& = & \frac{1}{4}\left(
\begin{array}{cccc}
  \eta ^2 \left(1+c_3\right) & 0 & 0 &  \eta  \left(c_1-c_2\right) \\
 0 & - \eta ^2 \left(-1+c_3\right) &  \eta  \left(c_1+c_2\right) & 0 \\
 0 & \eta  \left(c_1+c_2\right) &  \left(2-\eta ^2-\eta ^2 c_3\right) & 0 \\
  \eta  \left(c_1-c_2\right) & 0 & 0 &  \left(2-\eta ^2+\eta ^2 c_3\right)
\end{array}
\right).\label{eq:amplitude}
\end{eqnarray*}
\end{widetext}
One can readily find that the state ~$\varepsilon _a ( \rho )$ has the same form with state in Eq.(\ref{eq:xstates4})
\begin{equation*}
\varepsilon _a ( \rho ) = \frac{1}{4}\Big(I \otimes I + r(t)\sigma_3 \otimes I + s(t)I \otimes  \sigma_3 + \sum_{i=1}^3 c_i(t) \sigma_i \otimes \sigma_i \Big),
\label{eq:amplitude1}
\end{equation*}
here ~$ r(t)=\eta^2-1,s(t)=0,c_1(t)=\eta c_1,c_2(t)=\eta c_2,c_3(t)=\eta^2 c_3$, then by the formula given by \cite{PhysRevA.83.022321}, we have that
\begin{eqnarray*}
S_1 & = & 1+ \frac{1}{2} f (\eta^2-1 + \eta^2 c_3) + \frac{1}{2} f (\eta^2-1 - \eta^2 c_3)\label{s1amplitude},\nonumber\\
S_2 & = & 1+ f(\sqrt{(\eta^2-1)^2 + (\eta c_1)^2})\label{s2amplitude},\nonumber\\
S_3 & = & 1+ f(\sqrt{(\eta^2-1)^2 + (\eta c_2)^2})\label{s3amplitude}.\nonumber\\
\end{eqnarray*}
We  find that it is always true that  ~$S_3 < S_2 \   (\text{or}\  S_2 < S_3)$  if  ~$c_2 > c_1\ (\text{or}\  c_1 > c_2)$, so we just need to compare ~$S_1$ with ~$S_3\  (\text{or}\ S_2) $ to obtain the minimum of ~$\{S_1,S_2,S_3\}$. We note, ~$S_3 \leqslant S_1 $, always happens when ~$ c_2 \geqslant c_3$,
in comparison, the minimum of ~$\{S_1,S_2,S_3\}$ transfers from  ~$S_3$ to ~$ S_1 $  when ~$ c_2 < c_3$. We show this in Fig.(\ref{amplitude4}). The sudden change of quantum discord  happens  when ~$ c_2 <  c_3$ in  the initial state state ~$\rho$ with this kind of noise environment.

\section{phase noise}\label{phasenoise}
\begin{figure}
  % Requires \usepackage{graphicx}
  \includegraphics[width=0.5\textwidth]{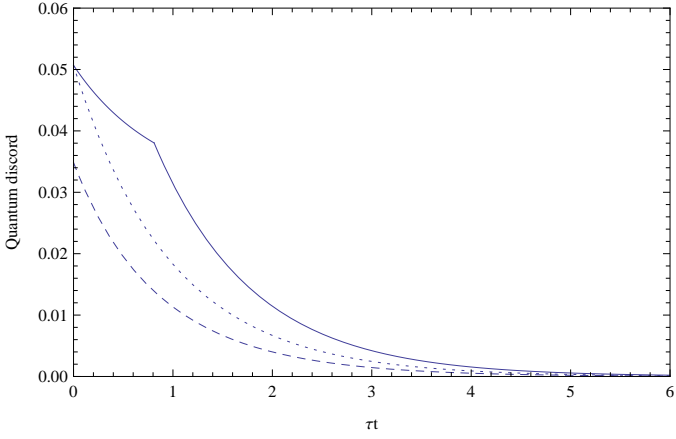}\\
  \caption{Quantum discord of ~$\rho$ under phase noise acting on the first qubit of the quantum system. (1)$c_1 = 0.1, c_2 = 0.2,c_3 = 0.3$ (dotted line). (2)$c_1 = 0.1,c_2 = 0.4,c_3 = 0.2$ (solid line). (3)$c_1 = c_2 =0.2,c_3 = 0$ (dashed line). The sudden change happens only at situation (2).}\label{xphase}
\end{figure}

 Phase noise or phase damping channel describes a quantum noise with loss of quantum phase information without loss of energy. The Kraus operators of this noise for single qubit are given by \cite{Nielsen,PhysRevA.78.022322}
 \begin{equation*}
K_{0}=
\begin{bmatrix}
1 & 0  \\  0 &  \gamma
\end{bmatrix}
, \ \
K_{1}=
\begin{bmatrix}
0 & 0  \\  0 &  \sqrt{1-\gamma^{2}}
\end{bmatrix},
\end{equation*}
where $\gamma =e^{-\frac{\tau t}{2}}$ and $\tau$ denotes transversal decay rate. The operators of phase noise acting on the first qubit of state $\rho$,
 so we have $K_{1p}=K_1 \otimes I_2,K_{2p}=K_2 \otimes I_2$, then one can find,

\begin{eqnarray}
&&\varepsilon_p(\rho)  =  K_{1p} \rho K_{1p}^\dagger + K_{2p}\rho K_{2p}^\dagger \nonumber\\
 & = & \frac{1}{4}\left(
\begin{array}{cccc}
 1+c_3 & 0 & 0 & \gamma  \left(c_1-c_2\right) \\
 0 & 1-c_3 & \gamma  \left(c_1+c_2\right) & 0 \\
 0 & \gamma  \left(c_1+c_2\right) & 1-c_3 & 0 \\
 \gamma  \left(c_1-c_2\right) & 0 & 0 & 1+c_3
\end{array}
\right)\nonumber\\
& = & \frac{1}{4}\Big(I+\gamma c_1\sigma_1\otimes \sigma_1+\gamma c_2\sigma_2\otimes \sigma_2+c_3\sigma_3\otimes \sigma_3\Big)\label{Eq:rhophase}
\end{eqnarray}
Comparing Eq. (\ref{Eq:rhophase}) with Eq. (\ref{eq:xstates4}), we can easily obtain the classical correlation and quantum correlation of state ~$\rho$ under phase noise acting on the first qubit,

\begin{equation}
\mathcal{C}(\varepsilon_p(\rho)) = \frac{1-\chi}{2} log_2 (1-\chi) + \frac{1+\chi}{2} log_2 (1+\chi),\label{phaseclassical}
\end{equation}
\begin{eqnarray}
&&\mathcal{Q}(\varepsilon_p(\rho)) = \mathcal{I}(\varepsilon_p(\rho))-\mathcal{C}(\varepsilon_p(\rho))\nonumber\\
            &=& 2+\sum_{i=1}^4 \lambda_i log_2 \lambda_i-\frac{1-\chi}{2} log_2 (1-\chi) - \frac{1+\chi}{2} log_2 (1+\chi),\nonumber\\\label{phasequantum}
\end{eqnarray}
where ~$\chi = \text{max} \{|\gamma c_1|,|\gamma c_2|,|c_3|\}$, and ~$\{\lambda_i\}$ are the eigenvalues of ~$\varepsilon_p(\rho)  $.
If ~$|c_3|\geqslant \text{max}\{|c_1|, |c_2|\}$,$\chi$ in Eq.(\ref{phaseclassical}) and Eq.(\ref{phasequantum}) will equal to  ~$|c_3|$, and the  classical correlation ~$\mathcal{C}(\varepsilon_p(\rho))$ remains unaffected, while the quantum correlation ~$\mathcal{Q}(\varepsilon _p (\rho))$ decays monotonically.
If  ~$\text{max}\{|c_1|, |c_2|\}\geqslant |c_3|$ and ~$|c_3|\neq 0$, the dynamics of  classical correlation ~$\mathcal{C}(\varepsilon_p(\rho))$ and quantum correlation ~$\mathcal{Q}(\varepsilon _p (\rho))$ have a sudden change at ~$t_0 = -\frac{2}{\tau} \log_2 |\frac{c_3}{\text{max}\{c_1,c_2\}}|$. In Fig. \ref{xphase}, we depict the dynamic of quantum discord of ~$\rho$ under single qubit phase noise with different $\{c_i\}$.
It is shown that sudden change of quantum discord can occur when phase noise act only on one part of a two-qubit quantum state.

\section{depolarizing noise}\label{deponoise}
\begin{figure}
  % Requires \usepackage{graphicx}
  \includegraphics[width=0.45\textwidth]{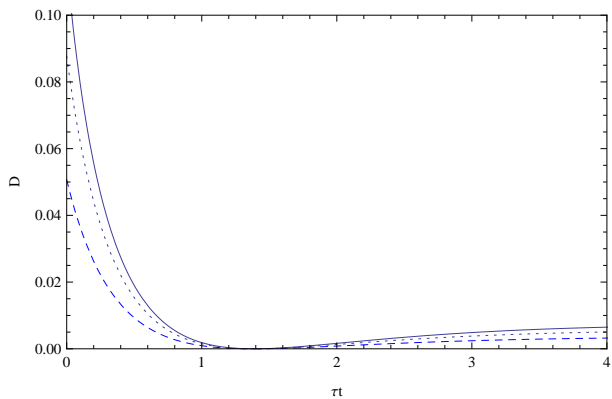}\\
  \caption{Quantum discord of  state ~$\rho$ under depolarizing noise with (1) $c_1=0.1,c_2=0.2,c_3=0.3$ (dashed line), (2)$c_1=0.1,c_2=0.4,c_3=0.3$ (solid line) and (3)$c_1=0.3,c_2=0.2,c_3=0.2$(dotted line) respectively. All of the three lines turn to 0 at ~$\Gamma t =log_2 4$.}\label{depolarizing3figure}
\end{figure}

\begin{figure}
  % Requires \usepackage{graphicx}
  \includegraphics[width=0.5\textwidth]{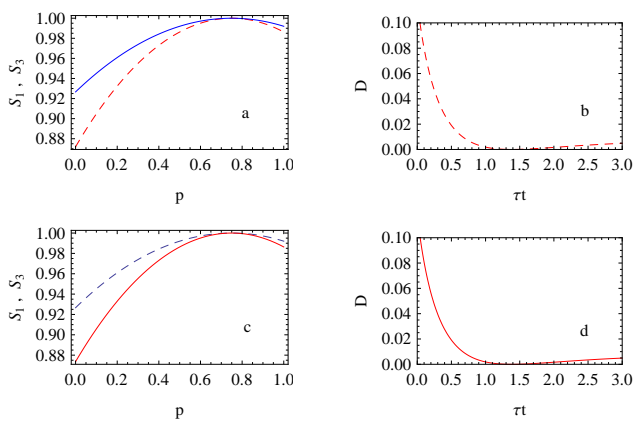}\\
  \caption{Depolarizing noise acting on state ~$\rho'$:~$S_1$ (dashed line) is always less than ~$S_3$(solid line) in (a)~$\{c_1=0.1,c_2=0.3,c_3=0.4,r=0.1,s = -0.01\}$ and the quantum discord in this situation is shown in (b); $S_3$ (solid line) is always less than $ S_1$ (dashed line) in (c)~$\{c_1=0.1,c_2=0.4,c_3=0.3,r=0.1,s = 0.01\}$, and we show quantum discord of this situation in (d).}\label{depolarizing5}
\end{figure}

The depolarizing noise is an important type of quantum noise that take a single qubit into completely mixed state $I/2$ with probability $p$ and leave itself untouched with probability $1-p$. The operators for single qubit depolarizing noise are given by \cite{Nielsen}
\begin{eqnarray*}
D_{1}&=& \sqrt{1-p}
\left(\begin{array}{rr}
1 & 0  \\  0 &  1
\end{array} \right)
, \ \
D_{2}= \sqrt{\frac{p}{3}}
\left(\begin{array}{rr}
0 & 1  \\  1 & 0
\end{array} \right), \nonumber \\
D_{3}&=&\sqrt{\frac{p}{3}}
\left(\begin{array}{rr}
0 & -i  \\  i &  0
\end{array} \right), \ \
D_{4}=\sqrt{\frac{p}{3}}
\left(\begin{array}{rr}
1 & 0  \\  0 &  -1
\end{array} \right).
\end{eqnarray*}
Where ~$ p = 1-e^{-\tau t}$,  then we have the operators ~$\{K_{id}\}$ acting on the first qubit of a composite system
$K_{1d}=D_{1}\otimes I_2,\ \ K_{2d}=D_{2}\otimes I_2, K_{3d}=D_{3}\otimes I_2,\ \ K_{2d}=D_{4}\otimes I_2.$
The two-qubit system under the depolarizing noise acting on the first qubit of quantum state $\rho$ is given as,
\begin{widetext}
\begin{eqnarray}
 \varepsilon_d(\rho)&=&\sum_{i=1}^4 \, K_{id}\,\rho\,K_{id}^\dagger \nonumber \\
&=& \frac{1}{4}\left(
\begin{array}{cccc}
 1+\left(1-\frac{4 p}{3}\right) c_3 & 0 & 0 & \left(1-\frac{4 p}{3}\right) \left(c_1-c_2\right) \\
 0 & 1-\left(1-\frac{4 p}{3}\right) c_3 & \left(1-\frac{4 p}{3}\right) \left(c_1+c_2\right) & 0 \\
 0 & \left(1-\frac{4 p}{3}\right) \left(c_1+c_2\right) & 1-\left(1-\frac{4 p}{3}\right) c_3 & 0 \\
 \left(1-\frac{4 p}{3}\right) \left(c_1-c_2\right) & 0 & 0 & 1+\left(1-\frac{4 p}{3}\right) c_3
\end{array}
\right).\label{statedepolarizing}
\end{eqnarray}
\end{widetext}
Comparing  Eq.(\ref{statedepolarizing}) with Eq.(\ref{eq:xstates4}), we obtain that
\begin{equation}
\varepsilon _d ( \rho ) = \frac{1}{4}\Big(I  \otimes I + r(t)\sigma_3 \otimes I + s(t)I \otimes  \sigma_3 + \sum_{i=1}^3 c_i(t) \sigma_i \otimes \sigma_i \Big),
\label{eq:depolarizing}
\end{equation}
here ~$c_1(t)=\left(1-\frac{4 p}{3}\right)c_1, c_2(t)=\left(1-\frac{4 p}{3}\right)c_2, c_3(t)=\left(1-\frac{4 p}{3}\right)c_3$. We find that the maximum of ~$\{c_i(t)\}$ is always decided by the maximum of $\{c_i\}$ which is fixed and is independent of time,
thus there is no sudden change of  decay rate for  the classical and quantum correlations of state $\rho$ under depolarizing noise. We show this in Fig.(\ref{depolarizing3figure}). Quantum discord decays to 0 at  $\Gamma t =log_2 4$, and this means state ~$\rho$ has turned to
a completely mixed state.
We have also consider  the case that state $\rho'$ is  under the depolarizing noise,
and we find that the minimum of $\{S_i\}$ does not change from $S_1$(or $S_3$)  to $S_3$ (or $S_1$) under this kind of noise.
Therefore, there is no sudden change for quantum discord of state $\rho'$ under depolarizing noise, see Fig.(\ref{depolarizing5}).

\section{conclusion}\label{conclusi}
We have studied the quantum discord of the X class of quantum states under three different kinds of noise such as amplitude, dephasing and depolarizing in this article. We  show that composite noise is not the necessary condition for the occurrence of sudden change of quantum discord and the sudden change can happen when the  quantum noise  act only  on one qubit of the quantum system. Especially, the classical correlation can remain unaffected under the phase noise acting only on the first qubit of state ~$\rho$ just like it occurs when ~$\rho$ is exposed into a composite noise environment \cite{PhysRevA.80.044102}.
On the other hand, we should note that
the single noise acting on only one of the qubits of an two-qubit quantum state is not sufficient for the happening of sudden change of quantum discord, the suitable state is also needed.

The properties of various quantum correlations are studied generally for different situations with both coherent and individual operations.
Quantum discord can demonstrate some special phenomena which can happen with only one side of quantum operation.
The results in this paper provide more evidences that quantum discord is different from the quantum entanglement.
Experimentally, the bipartite qubits system with various one side quantum channels can be realized in
optical, nuclear magnetic resonance, superconducting qubits and solid states such as nitrogen-vacancy diamond
systems. The behavior of sudden change in quantum discord should be observed explicitly.

We notice a related paper \cite{PhysRevLett.109.190402} 
in which the behaviors of quantum correlations under phase damping and amplitude damping channels 
acting only on the apparatus are considered. And the sudden change of discord is also observed which
confirms our conclusion. On the other hand, 
there are some differences  between our results and theirs. 
In \cite{PhysRevLett.109.190402}, the maximal classical correlation, which appears as one part in 
definition of quantum discord, can be achieved by two different measurement projectors, i.e., 
in  the $\sigma_z$ basis or in the  $\sigma_x$ basis. In comparison, our results need
to consider the minimum of $S_1, S_2, S_3$ as shown in Eq. (\ref{liboresult}) introduced in
 Ref.\cite{PhysRevA.83.022321},
so the explicit analytic expressions for the dynamics
of quantum correlation under different kinds of noise can be found.

\emph{Acknowledgements:} This work is supported by ``973'' program (2010CB922904) and NSFC (11175248, 10875060).
We would like to thank F. F. Fanchini for pointing out Ref.\cite{PhysRevLett.109.190402} for us.


\begin{thebibliography}{38}
\expandafter\ifx\csname natexlab\endcsname\relax\def\natexlab#1{#1}\fi
\expandafter\ifx\csname bibnamefont\endcsname\relax
  \def\bibnamefont#1{#1}\fi
\expandafter\ifx\csname bibfnamefont\endcsname\relax
  \def\bibfnamefont#1{#1}\fi
\expandafter\ifx\csname citenamefont\endcsname\relax
  \def\citenamefont#1{#1}\fi
\expandafter\ifx\csname url\endcsname\relax
  \def\url#1{\texttt{#1}}\fi
\expandafter\ifx\csname urlprefix\endcsname\relax\def\urlprefix{URL }\fi
\providecommand{\bibinfo}[2]{#2}
\providecommand{\eprint}[2][]{\url{#2}}

\bibitem[{\citenamefont{Schr\"{o}dinger}(1935)}]{Naturwissenschaften.23.807-812}
\bibinfo{author}{\bibfnamefont{E.}~\bibnamefont{Schr\"{o}dinger}},
  \bibinfo{journal}{Naturwissenschaften} \textbf{\bibinfo{volume}{23}},
  \bibinfo{pages}{807} (\bibinfo{year}{1935}).

\bibitem[{\citenamefont{Einstein et~al.}(1935)\citenamefont{Einstein, Podolsky,
  and Rosen}}]{PhysRev.47.777}
\bibinfo{author}{\bibfnamefont{A.}~\bibnamefont{Einstein}},
  \bibinfo{author}{\bibfnamefont{B.}~\bibnamefont{Podolsky}}, \bibnamefont{and}
  \bibinfo{author}{\bibfnamefont{N.}~\bibnamefont{Rosen}},
  \bibinfo{journal}{Phys. Rev.} \textbf{\bibinfo{volume}{47}},
  \bibinfo{pages}{777} (\bibinfo{year}{1935}).

\bibitem[{\citenamefont{Horodecki et~al.}(2009)\citenamefont{Horodecki,
  Horodecki, Horodecki, and Horodecki}}]{RevModPhys.81.865}
\bibinfo{author}{\bibfnamefont{R.}~\bibnamefont{Horodecki}},
  \bibinfo{author}{\bibfnamefont{P.}~\bibnamefont{Horodecki}},
  \bibinfo{author}{\bibfnamefont{M.}~\bibnamefont{Horodecki}},
  \bibnamefont{and}
  \bibinfo{author}{\bibfnamefont{K.}~\bibnamefont{Horodecki}},
  \bibinfo{journal}{Rev. Mod. Phys.} \textbf{\bibinfo{volume}{81}},
  \bibinfo{pages}{865} (\bibinfo{year}{2009}).

\bibitem[{\citenamefont{Michael.A.Nielsen}(2000)}]{Nielsen}
\bibinfo{author}{\bibfnamefont{I.~L.} \bibnamefont{Michael.A.Nielsen}},
  \emph{\bibinfo{title}{Quantum Computation and Quantum Information}}
  (\bibinfo{publisher}{Cambridge University Press}, \bibinfo{year}{2000}).

\bibitem[{\citenamefont{Modi et~al.}(2010)\citenamefont{Modi, Paterek, Son,
  Vedral, and Williamson}}]{PhysRevLett.104.080501}
\bibinfo{author}{\bibfnamefont{K.}~\bibnamefont{Modi}},
  \bibinfo{author}{\bibfnamefont{T.}~\bibnamefont{Paterek}},
  \bibinfo{author}{\bibfnamefont{W.}~\bibnamefont{Son}},
  \bibinfo{author}{\bibfnamefont{V.}~\bibnamefont{Vedral}}, \bibnamefont{and}
  \bibinfo{author}{\bibfnamefont{M.}~\bibnamefont{Williamson}},
  \bibinfo{journal}{Phys. Rev. Lett.} \textbf{\bibinfo{volume}{104}},
  \bibinfo{pages}{080501} (\bibinfo{year}{2010}).

\bibitem[{\citenamefont{Ollivier and Zurek}(2001)}]{PhysRevLett.88.017901}
\bibinfo{author}{\bibfnamefont{H.}~\bibnamefont{Ollivier}} \bibnamefont{and}
  \bibinfo{author}{\bibfnamefont{W.~H.} \bibnamefont{Zurek}},
  \bibinfo{journal}{Phys. Rev. Lett.} \textbf{\bibinfo{volume}{88}},
  \bibinfo{pages}{017901} (\bibinfo{year}{2001}).

\bibitem[{\citenamefont{Modi et~al.}(2012)\citenamefont{Modi, Brodutch, Cable,
  Paterek, and Vedral}}]{RevModPhys.84.1655}
\bibinfo{author}{\bibfnamefont{K.}~\bibnamefont{Modi}},
  \bibinfo{author}{\bibfnamefont{A.}~\bibnamefont{Brodutch}},
  \bibinfo{author}{\bibfnamefont{H.}~\bibnamefont{Cable}},
  \bibinfo{author}{\bibfnamefont{T.}~\bibnamefont{Paterek}}, \bibnamefont{and}
  \bibinfo{author}{\bibfnamefont{V.}~\bibnamefont{Vedral}},
  \bibinfo{journal}{Rev. Mod. Phys.} \textbf{\bibinfo{volume}{84}},
  \bibinfo{pages}{1655} (\bibinfo{year}{2012}).

\bibitem[{\citenamefont{Henderson and Vedra}(2001)}]{J.Phys.A.Math.Gen.34.6899}
\bibinfo{author}{\bibfnamefont{L.}~\bibnamefont{Henderson}} \bibnamefont{and}
  \bibinfo{author}{\bibfnamefont{V.}~\bibnamefont{Vedra}},
  \bibinfo{journal}{Journal of Physics A: Mathematical and General}
  \textbf{\bibinfo{volume}{34}}, \bibinfo{pages}{6899} (\bibinfo{year}{2001}).

\bibitem[{\citenamefont{Zurek}(2003)}]{PhysRevA.67.012320}
\bibinfo{author}{\bibfnamefont{W.~H.} \bibnamefont{Zurek}},
  \bibinfo{journal}{Phys. Rev. A} \textbf{\bibinfo{volume}{67}},
  \bibinfo{pages}{012320} (\bibinfo{year}{2003}).

\bibitem[{\citenamefont{Horodecki et~al.}(2005)\citenamefont{Horodecki,
  Horodecki, Horodecki, Oppenheim, Sen(De), Sen, and
  Synak-Radtke}}]{PhysRevA.71.062307}
\bibinfo{author}{\bibfnamefont{M.}~\bibnamefont{Horodecki}},
  \bibinfo{author}{\bibfnamefont{P.}~\bibnamefont{Horodecki}},
  \bibinfo{author}{\bibfnamefont{R.}~\bibnamefont{Horodecki}},
  \bibinfo{author}{\bibfnamefont{J.}~\bibnamefont{Oppenheim}},
  \bibinfo{author}{\bibfnamefont{A.}~\bibnamefont{Sen(De)}},
  \bibinfo{author}{\bibfnamefont{U.}~\bibnamefont{Sen}}, \bibnamefont{and}
  \bibinfo{author}{\bibfnamefont{B.}~\bibnamefont{Synak-Radtke}},
  \bibinfo{journal}{Phys. Rev. A} \textbf{\bibinfo{volume}{71}},
  \bibinfo{pages}{062307} (\bibinfo{year}{2005}).

\bibitem[{\citenamefont{Li and Luo}(2007)}]{PhysRevA.76.032327}
\bibinfo{author}{\bibfnamefont{N.}~\bibnamefont{Li}} \bibnamefont{and}
  \bibinfo{author}{\bibfnamefont{S.}~\bibnamefont{Luo}},
  \bibinfo{journal}{Phys. Rev. A} \textbf{\bibinfo{volume}{76}},
  \bibinfo{pages}{032327} (\bibinfo{year}{2007}).

\bibitem[{\citenamefont{Luo}(2008)}]{PhysRevA.77.042303}
\bibinfo{author}{\bibfnamefont{S.}~\bibnamefont{Luo}}, \bibinfo{journal}{Phys.
  Rev. A} \textbf{\bibinfo{volume}{77}}, \bibinfo{pages}{042303}
  (\bibinfo{year}{2008}).

\bibitem[{\citenamefont{Werlang et~al.}(2009)\citenamefont{Werlang, Souza,
  Fanchini, and Villas~Boas}}]{PhysRevA.80.024103}
\bibinfo{author}{\bibfnamefont{T.}~\bibnamefont{Werlang}},
  \bibinfo{author}{\bibfnamefont{S.}~\bibnamefont{Souza}},
  \bibinfo{author}{\bibfnamefont{F.~F.} \bibnamefont{Fanchini}},
  \bibnamefont{and} \bibinfo{author}{\bibfnamefont{C.~J.}
  \bibnamefont{Villas~Boas}}, \bibinfo{journal}{Phys. Rev. A}
  \textbf{\bibinfo{volume}{80}}, \bibinfo{pages}{024103}
  (\bibinfo{year}{2009}).

\bibitem[{\citenamefont{Sarandy}(2009)}]{PhysRevA.80.022108}
\bibinfo{author}{\bibfnamefont{M.~S.} \bibnamefont{Sarandy}},
  \bibinfo{journal}{Phys. Rev. A} \textbf{\bibinfo{volume}{80}},
  \bibinfo{pages}{022108} (\bibinfo{year}{2009}).

\bibitem[{\citenamefont{Maziero et~al.}(2009)\citenamefont{Maziero, C\'eleri,
  Serra, and Vedral}}]{PhysRevA.80.044102}
\bibinfo{author}{\bibfnamefont{J.}~\bibnamefont{Maziero}},
  \bibinfo{author}{\bibfnamefont{L.~C.} \bibnamefont{C\'eleri}},
  \bibinfo{author}{\bibfnamefont{R.~M.} \bibnamefont{Serra}}, \bibnamefont{and}
  \bibinfo{author}{\bibfnamefont{V.}~\bibnamefont{Vedral}},
  \bibinfo{journal}{Phys. Rev. A} \textbf{\bibinfo{volume}{80}},
  \bibinfo{pages}{044102} (\bibinfo{year}{2009}).

\bibitem[{\citenamefont{Datta}(2009)}]{PhysRevA.80.052304}
\bibinfo{author}{\bibfnamefont{A.}~\bibnamefont{Datta}},
  \bibinfo{journal}{Phys. Rev. A} \textbf{\bibinfo{volume}{80}},
  \bibinfo{pages}{052304} (\bibinfo{year}{2009}).

\bibitem[{\citenamefont{Ali et~al.}(2010{\natexlab{a}})\citenamefont{Ali, Rau,
  and Alber}}]{PhysRevA.81.042105}
\bibinfo{author}{\bibfnamefont{M.}~\bibnamefont{Ali}},
  \bibinfo{author}{\bibfnamefont{A.~R.~P.} \bibnamefont{Rau}},
  \bibnamefont{and} \bibinfo{author}{\bibfnamefont{G.}~\bibnamefont{Alber}},
  \bibinfo{journal}{Phys. Rev. A} \textbf{\bibinfo{volume}{81}},
  \bibinfo{pages}{042105} (\bibinfo{year}{2010}{\natexlab{a}});
  \textbf{\bibinfo{volume}{82}},
  \bibinfo{pages}{069902} (\bibinfo{year}{2010}{\natexlab{b}}).

\bibitem[{\citenamefont{Chen et~al.}(2011)\citenamefont{Chen, Zhang, Yu, Yi,
  and Oh}}]{PhysRevA.84.042313}
\bibinfo{author}{\bibfnamefont{Q.}~\bibnamefont{Chen}},
  \bibinfo{author}{\bibfnamefont{C.}~\bibnamefont{Zhang}},
  \bibinfo{author}{\bibfnamefont{S.~X.}~\bibnamefont{Yu}},
  \bibinfo{author}{\bibfnamefont{X.~X.} \bibnamefont{Yi}}, \bibnamefont{and}
  \bibinfo{author}{\bibfnamefont{C.~H.} \bibnamefont{Oh}},
  \bibinfo{journal}{Phys. Rev. A} \textbf{\bibinfo{volume}{84}},
  \bibinfo{pages}{042313} (\bibinfo{year}{2011}).

\bibitem[{\citenamefont{Li et~al.}(2011)\citenamefont{Li, Wang, and
  Fei}}]{PhysRevA.83.022321}
\bibinfo{author}{\bibfnamefont{B.}~\bibnamefont{Li}},
  \bibinfo{author}{\bibfnamefont{Z.-X.} \bibnamefont{Wang}}, \bibnamefont{and}
  \bibinfo{author}{\bibfnamefont{S.-M.} \bibnamefont{Fei}},
  \bibinfo{journal}{Phys. Rev. A} \textbf{\bibinfo{volume}{83}},
  \bibinfo{pages}{022321} (\bibinfo{year}{2011}).

\bibitem[{\citenamefont{Chitambar}(2012)}]{PhysRevA.86.032110}
\bibinfo{author}{\bibfnamefont{E.}~\bibnamefont{Chitambar}},
  \bibinfo{journal}{Phys. Rev. A} \textbf{\bibinfo{volume}{86}},
  \bibinfo{pages}{032110} (\bibinfo{year}{2012}).

\bibitem[{\citenamefont{Datta et~al.}(2008)\citenamefont{Datta, Shaji, and
  Caves}}]{PhysRevLett.100.050502}
\bibinfo{author}{\bibfnamefont{A.}~\bibnamefont{Datta}},
  \bibinfo{author}{\bibfnamefont{A.}~\bibnamefont{Shaji}}, \bibnamefont{and}
  \bibinfo{author}{\bibfnamefont{C.~M.} \bibnamefont{Caves}},
  \bibinfo{journal}{Phys. Rev. Lett.} \textbf{\bibinfo{volume}{100}},
  \bibinfo{pages}{050502} (\bibinfo{year}{2008}).

\bibitem[{\citenamefont{Piani et~al.}(2008)\citenamefont{Piani, Horodecki, and
  Horodecki}}]{PhysRevLett.100.090502}
\bibinfo{author}{\bibfnamefont{M.}~\bibnamefont{Piani}},
  \bibinfo{author}{\bibfnamefont{P.}~\bibnamefont{Horodecki}},
  \bibnamefont{and}
  \bibinfo{author}{\bibfnamefont{R.}~\bibnamefont{Horodecki}},
  \bibinfo{journal}{Phys. Rev. Lett.} \textbf{\bibinfo{volume}{100}},
  \bibinfo{pages}{090502} (\bibinfo{year}{2008}).

\bibitem[{\citenamefont{Piani et~al.}(2009)\citenamefont{Piani, Christandl,
  Mora, and Horodecki}}]{PhysRevLett.102.250503}
\bibinfo{author}{\bibfnamefont{M.}~\bibnamefont{Piani}},
  \bibinfo{author}{\bibfnamefont{M.}~\bibnamefont{Christandl}},
  \bibinfo{author}{\bibfnamefont{C.~E.} \bibnamefont{Mora}}, \bibnamefont{and}
  \bibinfo{author}{\bibfnamefont{P.}~\bibnamefont{Horodecki}},
  \bibinfo{journal}{Phys. Rev. Lett.} \textbf{\bibinfo{volume}{102}},
  \bibinfo{pages}{250503} (\bibinfo{year}{2009}).

\bibitem[{\citenamefont{Shabani and Lidar}(2009)}]{PhysRevLett.102.100402}
\bibinfo{author}{\bibfnamefont{A.}~\bibnamefont{Shabani}} \bibnamefont{and}
  \bibinfo{author}{\bibfnamefont{D.~A.} \bibnamefont{Lidar}},
  \bibinfo{journal}{Phys. Rev. Lett.} \textbf{\bibinfo{volume}{102}},
  \bibinfo{pages}{100402} (\bibinfo{year}{2009}).

\bibitem[{\citenamefont{Daki\ifmmode~\acute{c}\else \'{c}\fi{}
  et~al.}(2010)\citenamefont{Daki\ifmmode~\acute{c}\else \'{c}\fi{}, Vedral,
  and \ifmmode \check{C}\else~\v{C}\fi{}. Brukner}}]{PhysRevLett.105.190502}
\bibinfo{author}{\bibfnamefont{B.}~\bibnamefont{Daki\ifmmode~\acute{c}\else
  \'{c}\fi{}}}, \bibinfo{author}{\bibfnamefont{V.}~\bibnamefont{Vedral}},
  \bibnamefont{and} \bibinfo{author}{\bibnamefont{\ifmmode
  \check{C}\else~\v{C}\fi{}. Brukner}}, \bibinfo{journal}{Phys. Rev. Lett.}
  \textbf{\bibinfo{volume}{105}}, \bibinfo{pages}{190502}
  (\bibinfo{year}{2010}).

\bibitem[{\citenamefont{Mazzola et~al.}(2010)\citenamefont{Mazzola, Piilo, and
  Maniscalco}}]{PhysRevLett.104.200401}
\bibinfo{author}{\bibfnamefont{L.}~\bibnamefont{Mazzola}},
  \bibinfo{author}{\bibfnamefont{J.}~\bibnamefont{Piilo}}, \bibnamefont{and}
  \bibinfo{author}{\bibfnamefont{S.}~\bibnamefont{Maniscalco}},
  \bibinfo{journal}{Phys. Rev. Lett.} \textbf{\bibinfo{volume}{104}},
  \bibinfo{pages}{200401} (\bibinfo{year}{2010}).

\bibitem[{\citenamefont{Lo~Franco et~al.}(2012)\citenamefont{Lo~Franco,
  Bellomo, Andersson, and Compagno}}]{PhysRevA.85.032318}
\bibinfo{author}{\bibfnamefont{R.}~\bibnamefont{Lo~Franco}},
  \bibinfo{author}{\bibfnamefont{B.}~\bibnamefont{Bellomo}},
  \bibinfo{author}{\bibfnamefont{E.}~\bibnamefont{Andersson}},
  \bibnamefont{and} \bibinfo{author}{\bibfnamefont{G.}~\bibnamefont{Compagno}},
  \bibinfo{journal}{Phys. Rev. A} \textbf{\bibinfo{volume}{85}},
  \bibinfo{pages}{032318} (\bibinfo{year}{2012}).

\bibitem[{\citenamefont{Bellomo et~al.}(2012)\citenamefont{Bellomo, Lo~Franco,
  and Compagno}}]{PhysRevA.86.012312}
\bibinfo{author}{\bibfnamefont{B.}~\bibnamefont{Bellomo}},
  \bibinfo{author}{\bibfnamefont{R.}~\bibnamefont{Lo~Franco}},
  \bibnamefont{and} \bibinfo{author}{\bibfnamefont{G.}~\bibnamefont{Compagno}},
  \bibinfo{journal}{Phys. Rev. A} \textbf{\bibinfo{volume}{86}},
  \bibinfo{pages}{012312} (\bibinfo{year}{2012}).

\bibitem[{\citenamefont{G.~Karpat}(2012)}]{arXiv:1208.5705}
\bibinfo{author}{\bibnamefont{G.~Karpat}
\bibnamefont{and}
\bibfnamefont{Z.~Gedik} },
  \bibinfo{journal}{arXiv:1208.5705}  (\bibinfo{year}{2012}).


\bibitem[{\citenamefont{Lanyon et~al.}(2008)\citenamefont{Lanyon, Barbieri,
  Almeida, and White}}]{PhysRevLett.101.200501}
\bibinfo{author}{\bibfnamefont{B.~P.} \bibnamefont{Lanyon}},
  \bibinfo{author}{\bibfnamefont{M.}~\bibnamefont{Barbieri}},
  \bibinfo{author}{\bibfnamefont{M.~P.} \bibnamefont{Almeida}},
  \bibnamefont{and} \bibinfo{author}{\bibfnamefont{A.~G.} \bibnamefont{White}},
  \bibinfo{journal}{Phys. Rev. Lett.} \textbf{\bibinfo{volume}{101}},
  \bibinfo{pages}{200501} (\bibinfo{year}{2008}).


\bibitem[{\citenamefont{Roa et~al.}(2011)\citenamefont{Roa, Retamal, and
  Alid-Vaccarezza}}]{PhysRevLett.107.080401}
\bibinfo{author}{\bibfnamefont{L.}~\bibnamefont{Roa}},
  \bibinfo{author}{\bibfnamefont{J.~C.} \bibnamefont{Retamal}},
  \bibnamefont{and}
  \bibinfo{author}{\bibfnamefont{M.}~\bibnamefont{Alid-Vaccarezza}},
  \bibinfo{journal}{Phys. Rev. Lett.} \textbf{\bibinfo{volume}{107}},
  \bibinfo{pages}{080401} (\bibinfo{year}{2011}).

\bibitem[{\citenamefont{Li et~al.}(2012)\citenamefont{Li, Fei, Wang, and
  Fan}}]{PhysRevA.85.022328}
\bibinfo{author}{\bibfnamefont{B.}~\bibnamefont{Li}},
  \bibinfo{author}{\bibfnamefont{S.-M.} \bibnamefont{Fei}},
  \bibinfo{author}{\bibfnamefont{Z.-X.} \bibnamefont{Wang}}, \bibnamefont{and}
  \bibinfo{author}{\bibfnamefont{H.}~\bibnamefont{Fan}},
  \bibinfo{journal}{Phys. Rev. A} \textbf{\bibinfo{volume}{85}},
  \bibinfo{pages}{022328} (\bibinfo{year}{2012}).

\bibitem[{\citenamefont{Streltsov
  et~al.}(2011{\natexlab{a}})\citenamefont{Streltsov, Kampermann, and
  Bru\ss{}}}]{PhysRevLett.106.160401}
\bibinfo{author}{\bibfnamefont{A.}~\bibnamefont{Streltsov}},
  \bibinfo{author}{\bibfnamefont{H.}~\bibnamefont{Kampermann}},
  \bibnamefont{and} \bibinfo{author}{\bibfnamefont{D.}~\bibnamefont{Bru\ss{}}},
  \bibinfo{journal}{Phys. Rev. Lett.} \textbf{\bibinfo{volume}{106}},
  \bibinfo{pages}{160401} (\bibinfo{year}{2011}{\natexlab{a}}).

\bibitem[{\citenamefont{Koashi and Winter}(2004)}]{PhysRevA.69.022309}
\bibinfo{author}{\bibfnamefont{M.}~\bibnamefont{Koashi}} \bibnamefont{and}
  \bibinfo{author}{\bibfnamefont{A.}~\bibnamefont{Winter}},
  \bibinfo{journal}{Phys. Rev. A} \textbf{\bibinfo{volume}{69}},
  \bibinfo{pages}{022309} (\bibinfo{year}{2004}).

\bibitem[{\citenamefont{Streltsov
  et~al.}(2011{\natexlab{b}})\citenamefont{Streltsov, Kampermann, and
  Bru\ss{}}}]{PhysRevLett.107.170502}
\bibinfo{author}{\bibfnamefont{A.}~\bibnamefont{Streltsov}},
  \bibinfo{author}{\bibfnamefont{H.}~\bibnamefont{Kampermann}},
  \bibnamefont{and} \bibinfo{author}{\bibfnamefont{D.}~\bibnamefont{Bru\ss{}}},
  \bibinfo{journal}{Phys. Rev. Lett.} \textbf{\bibinfo{volume}{107}},
  \bibinfo{pages}{170502} (\bibinfo{year}{2011}{\natexlab{b}}).

\bibitem[{\citenamefont{Hu et~al.}(2012)\citenamefont{Hu, Fan, Zhou, and
  Liu}}]{PhysRevA.85.032102}
\bibinfo{author}{\bibfnamefont{X.}~\bibnamefont{Hu}},
  \bibinfo{author}{\bibfnamefont{H.}~\bibnamefont{Fan}},
  \bibinfo{author}{\bibfnamefont{D.~L.} \bibnamefont{Zhou}}, \bibnamefont{and}
  \bibinfo{author}{\bibfnamefont{W.-M.} \bibnamefont{Liu}},
  \bibinfo{journal}{Phys. Rev. A} \textbf{\bibinfo{volume}{85}},
  \bibinfo{pages}{032102} (\bibinfo{year}{2012}).

\bibitem[{\citenamefont{Yu and Eberly}(2004)}]{PhysRevLett.93.140404}
\bibinfo{author}{\bibfnamefont{T.}~\bibnamefont{Yu}} \bibnamefont{and}
  \bibinfo{author}{\bibfnamefont{J.~H.} \bibnamefont{Eberly}},
  \bibinfo{journal}{Phys. Rev. Lett.} \textbf{\bibinfo{volume}{93}},
  \bibinfo{pages}{140404} (\bibinfo{year}{2004}).

\bibitem[{\citenamefont{Yu and Eberly}(2006)}]{PhysRevLett.97.140403}
\bibinfo{author}{\bibfnamefont{T.}~\bibnamefont{Yu}} \bibnamefont{and}
  \bibinfo{author}{\bibfnamefont{J.~H.} \bibnamefont{Eberly}},
  \bibinfo{journal}{Phys. Rev. Lett.} \textbf{\bibinfo{volume}{97}},
  \bibinfo{pages}{140403} (\bibinfo{year}{2006}).

\bibitem[{\citenamefont{Yashodamma and Sudha}(2012)}]{arXiv:12103216v2}
\bibinfo{author}{\bibfnamefont{K.}~\bibfnamefont{O.}~\bibnamefont{Yashodamma}} \bibnamefont{and}
  \bibinfo{author}{\bibnamefont{Sudha}}, \bibinfo{journal}{arXiv:1210.3216}
  (\bibinfo{year}{2012}).

\bibitem[{\citenamefont{Ferraro et~al.}(2010)\citenamefont{Ferraro, Aolita,
  Cavalcanti, Cucchietti, and Ac\'{i}n}}]{PhysRevA.81.052318}
\bibinfo{author}{\bibfnamefont{A.}~\bibnamefont{Ferraro}},
  \bibinfo{author}{\bibfnamefont{L.}~\bibnamefont{Aolita}},
  \bibinfo{author}{\bibfnamefont{D.}~\bibnamefont{Cavalcanti}},
  \bibinfo{author}{\bibfnamefont{F.~M.} \bibnamefont{Cucchietti}},
  \bibnamefont{and} \bibinfo{author}{\bibfnamefont{A.}~\bibnamefont{Ac\'{i}n}},
  \bibinfo{journal}{Phys. Rev. A} \textbf{\bibinfo{volume}{81}},
  \bibinfo{pages}{052318} (\bibinfo{year}{2010}).




\bibitem[{\citenamefont{Salles et~al.}(2008)\citenamefont{Salles, de~Melo,
  Almeida, Hor-Meyll, Walborn, Souto~Ribeiro, and
  Davidovich}}]{PhysRevA.78.022322}
\bibinfo{author}{\bibfnamefont{A.}~\bibnamefont{Salles}},
  \bibinfo{author}{\bibfnamefont{F.}~\bibnamefont{de~Melo}},
  \bibinfo{author}{\bibfnamefont{M.~P.} \bibnamefont{Almeida}},
  \bibinfo{author}{\bibfnamefont{M.}~\bibnamefont{Hor-Meyll}},
  \bibinfo{author}{\bibfnamefont{S.~P.} \bibnamefont{Walborn}},
  \bibinfo{author}{\bibfnamefont{P.~H.} \bibnamefont{Souto~Ribeiro}},
  \bibnamefont{and}
  \bibinfo{author}{\bibfnamefont{L.}~\bibnamefont{Davidovich}},
  \bibinfo{journal}{Phys. Rev. A} \textbf{\bibinfo{volume}{78}},
  \bibinfo{pages}{022322} (\bibinfo{year}{2008}).

\bibitem[{\citenamefont{Cornelio et~al.}(2012)\citenamefont{Cornelio,
  Far\'{\i}as, Fanchini, Frerot, Aguilar, Hor-Meyll, de~Oliveira, Walborn,
  Caldeira, and Ribeiro}}]{PhysRevLett.109.190402}
\bibinfo{author}{\bibfnamefont{M.~F.} \bibnamefont{Cornelio}},
  \bibinfo{author}{\bibfnamefont{O.~J.} \bibnamefont{Far\'{\i}as}},
  \bibinfo{author}{\bibfnamefont{F.~F.} \bibnamefont{Fanchini}},
  \bibinfo{author}{\bibfnamefont{I.}~\bibnamefont{Frerot}},
  \bibinfo{author}{\bibfnamefont{G.~H.} \bibnamefont{Aguilar}},
  \bibinfo{author}{\bibfnamefont{M.~O.} \bibnamefont{Hor-Meyll}},
  \bibinfo{author}{\bibfnamefont{M.~C.} \bibnamefont{de~Oliveira}},
  \bibinfo{author}{\bibfnamefont{S.~P.} \bibnamefont{Walborn}},
  \bibinfo{author}{\bibfnamefont{A.~O.} \bibnamefont{Caldeira}},
  \bibnamefont{and} \bibinfo{author}{\bibfnamefont{P.~H.~S.}
  \bibnamefont{Ribeiro}}, \bibinfo{journal}{Phys. Rev. Lett.}
  \textbf{\bibinfo{volume}{109}}, \bibinfo{pages}{190402}
  (\bibinfo{year}{2012}).

\end{thebibliography}
\end{document}